\documentclass[12pt,nohyper,notoc]{JHEP}

\usepackage{amsmath2000,euscript,array,amssymb,cite} 

\def\N{{\cal N}}

\def\SU{\text{SU}}
\def\U{\text{U}}
\def\GL{\text{GL}}
\def\SL{\text{SL}}

\def\Dbarslash{\,\,{\raise.15ex\hbox{/}\mkern-12mu {\bar\D}}}
\def\Dslash{\,\,{\raise.15ex\hbox{/}\mkern-12mu \D}}
\def\delslash{\,\,{\raise.15ex\hbox{/}\mkern-9mu \partial}}
\def\delbarslash{\,\,{\raise.15ex\hbox{/}\mkern-9mu {\bar\partial}}}

\def\S{{\EuScript S}}
\def\J{{\EuScript J}}

\newcommand{\MAT}[1]{\begin{pmatrix} #1\end{pmatrix}}
\newcommand{\EQ}[1]{\begin{equation} #1 \end{equation}}
\newcommand{\AL}[1]{\begin{subequations}\begin{align} #1
\end{align}\end{subequations}}
\newcommand{\SP}[1]{\begin{equation}\begin{split} #1 \end{split}\end{equation}}

\title{The Phase Structure of Mass-Deformed SU(2)$\times$SU(2)
Quiver Theory}

\author{Timothy J.~Hollowood and Tom~Kingaby\\
Department of Physics, University of Wales Swansea,
Swansea, SA2 8PP, UK\\
E-mail: {\tt t.hollowood@swan.ac.uk}, {\tt pytk@swan.ac.uk}}

\abstract{The phase structure of the finite $\SU(2)\times\SU(2)$ theory
with $\N=2$ supersymmetry, broken to $\N=1$ by mass terms for the
adjoint-valued chiral multiplets, is determined exactly by compactifying the
theory on a circle of finite radius. The exact low-energy 
superpotential is constructed by identifying it as a linear
combination of the Hamiltonians of a certain symplectic reduction of 
the spin generalized elliptic Calogero-Moser integrable system. 
It is shown that the
theory has four confining, two Higgs and two massless Coulomb vacua
which agrees with a simple analysis of the tree-level
superpotential of the four-dimensional theory. In each vacuum, we
calculate all the condensates of the adjoint-valued scalars.
}

\keywords{}
\preprint{{\tt hep-th/0210096}\\SWAT-354}
\begin{document}

\section{Introduction}

This paper is concerned with the problem of calculating the phase
structure of certain finite $\N=2$ theories perturbed by mass terms
to $\N=1$. The main paradigm for this is the $\N=4$ theory perturbed by
mass terms to $\N=1$ (which one can think of 
passing through $\N=2$ on the way): the
so-called $\N=1^*$ theory. Let us describe this case in more detail.
Gauge theory with $\N=4$ supersymmetry is a finite theory for which
$S$-duality---actually $\SL(2,{\mathbb Z})$---is an exact symmetry. 
However, once broken to $\N=1$ by adding mass terms for the three adjoint
chiral multiplets, the duality is broken: instead of being an
exact symmetry it now relates different vacua of the theory 
\cite{donwitt,Nick}. For
instance, the weakly-coupled Higgs vacuum is related to one of 
the strongly-coupled
confining vacua by $\tau\to-1/\tau$, where $\tau$ is the usual
complexified coupling of the theory. 
More precisely, we know on the basis of semi-classical
reasoning that, for gauge group $\SU(N)$,
the vacua are associated to the partitions of $N$. Furthermore, those
with a mass gap are associated to the subset of 
equipartitions: $N=p\cdot q$. In
these vacua there is an unbroken $\SU(p)$ gauge symmetry and hence
using standard reasoning based on the Witten Index 
there should be an additional degeneracy of $p$. Consequently
the total number of massive vacua is equal to $\sum_{p|N}p$,
a sum over the integer divisors of $N$. In particular, the Higgs vacua
corresponds to $p=1$ and the $N$ confining vacua to $p=N$. These vacua 
form a finite-dimensional representation of $\SL(2,{\mathbb Z})$.

One way to investigate the vacuum structure of the mass-deformed
theory is to realize the mass deformation in a two-stage
process: first breaking to $\N=2$ with a massive adjoint
hypermultiplet. The Coulomb branch of the $\N=2$ theory is described in
the by-now standard way by a Seiberg-Witten curve \cite{Seiberg:1994rs}. 
Further breaking to $\N=1$ can be
understood as a perturbation which lifts most of the Coulomb branch to
leave the vacua of the $\N=1$ theory. In principle, the Seiberg-Witten curve
can be used to find the vacua and all the condensates of lowest
component chiral superfields in
each vacuum. In particular, the massive vacua are associated to points on
the Coulomb branch for which the associated Seiberg-Witten curve $\Sigma$
undergoes maximal degeneration. Since the curve $\Sigma$ is an
$N$-sheeted cover of the underlying torus with complex structure $\tau$
\cite{donwitt} the maximal degeneration involves
unbranched (unramified) $N$-fold covers of the torus. It is known
in the $\N=4$ case that the Seiberg-Witten curve is the spectral curve
of the elliptic Calogero-Moser integrable system \cite{Martinec:1995qn}.

An alternative and more direct approach for which the integrable
system plays a central r\^ole 
involves compactifying the four-dimensional theory
to three dimensions on a circle of finite radius 
\cite{Nick,Kapustin:1998xn}. 
In three dimensions, the $\N=2$ theory has a Coulomb branch of
twice the dimension of the four-dimensional theory due to the Wilson
lines and dual photons of the unbroken abelian gauge group. What is
particularly nice about this, is that the integrable dynamical system mentioned
above now stands centre stage
since the larger Coulomb branch is nothing but its
(complexified) phase space. In
contrast, the Coulomb branch of the four-dimensional theory only
corresponds to the action variables alone. In other words, the Wilson
lines and dual photons supply the missing angle variables. Further
soft breaking to $\N=1$ is realized by adding a superpotential which is
simply one of the action variables of the integrable system. The
superpotential calculated in this way 
is actually exact, {\it i.e.\/}~includes all the
quantum corrections. One way to see this is to interpret the whole
set-up in terms of the mirror map in three dimensions: in this case
the integrable system arises as the Higgs branch of the
mirror-dual theory in the form of a Hitchin system 
which, as such, is not subject to quantum corrections 
\cite{Kapustin:1998xn}. The superpotential 
is also independent of the compactification radius and so
vacua and condensates extracted from it are also valid in the
decoupling limit. 

In this picture, the vacua of the theory are identified with the critical
points of the exact superpotential and, since the latter is a
Hamiltonian of the integrable system, this means they are associated to
equilibrium positions for the evolution, or flow, 
generated by that Hamiltonian:
\begin{equation*}
\text{Vacuum}\qquad\longleftrightarrow\qquad\text{equilibrium position of a
given flow}
\end{equation*}
Massive vacua are
special in that they are equilibrium positions for {\it any\/} choice
of Hamiltonian in 
the space of action variables:
\begin{equation*}
\text{Massive Vacuum}\qquad\longleftrightarrow\qquad\text{equilibrium
position for all the flows}
\end{equation*}
Formally this follows from the
following line of reasoning. The
angle variables take values in the Jacobian of the spectral curve of
the integrable system (in this case the Seiberg-Witten
curve) $\Sigma$ and upon maximal degeneration of $\Sigma$ all the angle
variables are frozen to a point and nothing moves. More concretely, 
this fact was also
proved directly in the case of the ${\cal N}=1^*$ theory 
in the Appendix of \cite{Dorey:2002tj}. All the massive vacua
have been found in this case by extremizing the superpotential. For the
massless vacua the situation is not so well understood: although a
complete picture is available for $\SU(3)$, for $N>3$ there are only 
partial results \cite{Nick,oferandus}.

In \cite{DHK1,Hollowood:2002ax} 
the whole picture described above was generalized to
certain finite $\N=2$ theories, the so-called ``quiver models'', which
arise from certain brane configurations in string theory. These
theories have product $\SU(N)^k$ gauge symmetry. For these theories
there is also an underlying integrable system arising as a Hitchin system
which was identified as the spin-generalized elliptic 
Calogero-Moser system developed in 
Refs.~\cite{GH,Krichever:1994vg,Nekrasov:1996nq}. Once again the Coulomb
branch of the compactified theory is identified with the phase space of
the (complexified) integrable system and the exact
superpotential describing the breaking to $\N=1$ is one of the Hamiltonians. 

A complete picture of
all the massive vacua was found in \cite{DHK1} generalizing the situation
in the $\N=1^*$ theory. However, as in the $\N=1^*$ theory, 
the situation with the massless vacua is
not understood. This provides the motivation for the present
work. In it we shall investigate the simplest quiver model with
gauge group $\SU(2)\times\SU(2)$. In this case, we will be able to
find the complete vacuum structure explicitly including both the
previously known massive but now also the massless vacua.

\section{The phase structure via semi-classical reasoning}

In this section we shall infer the phase structure of the theory 
by investigating the tree-level superpotential.

The $\SU(2)\times\SU(2)$ $\N=2$ supersymmetric quiver theory is an
example of the more general $\SU(N)^k$ theories which we now define.
The field content consists of (i) for each $\SU(N)$ factor an $\N=1$
vector multiplet and adjoint-valued chiral multiplet $\Phi_i$, $i=1,\ldots,k$, 
and (ii) chiral
multiplets $Q_i$,
$\tilde Q_i$, $i=1,\ldots,k$, in the $({\bf N},\bar{\bf N})$ and
$(\bar{\bf N},{\bf N})$ of 
$\SU(N)_i\times\SU(N)_{i+1}$, respectively. The tree-level 
superpotential, including the mass-deformation to $\N=1$, has the form \EQ{
W=\frac1{g^2}{\rm
Tr}\,\Big\{\Phi_iQ_i\tilde Q_i-Q_i\Phi_{i+1}\tilde Q_i+m_iQ_i\tilde
Q_i+\mu_i\Phi_i^2\Big\}\ ,
\label{supp} } 
where we assume that the labels are defined modulo
$N$. Here, $m_i$
are the $\N=2$
supersymmetry preserving masses of the hypermultiplets and $\mu_i$ are the
$\N=2\to\N=1$ breaking
masses of the adjoint chiral multiplets.

We can investigate the vacuum structure of the $\SU(2)\times\SU(2)$
theory by solving the $F$-flatness conditions
modulo complex gauge
transformations in the usual way. The analysis was done for the
massive vacua in the more general
setting of the
$\SU(N)^k$ theory in \cite{DHK1} and we can simply quote the
results in this case. The solutions for the massless vacua are new.

First of all, there are confining vacua for which $\Phi_i=Q_i=\tilde
Q_i=0$ and the gauge symmetry is
completely unbroken. We expect that the theory at 
low energy is pure $\N=1$ Yang-Mills
with gauge group
$\SU(2)\times\SU(2)$. Since each $\SU(2)$ factor is independent and 
each on its own yields two independent vacua, in all we expect four
confining vacua.

There are two Higgs vacua in which the gauge group is completely
broken. For the first
\SP{ \Phi_1&=\tfrac 12(m_1+m_2){\rm
diag}\big(1,-1\big)\ ,\qquad\Phi_2= \tfrac12(m_1-m_2){\rm diag}
\big(1,-1\big)\ ,\\
Q_1&=\MAT{0 & 0\\ 1 & 0}\ ,\quad \tilde Q_1=\MAT{0
& m_1(\mu_1+\mu_2)+m_2(\mu_1-\mu_2) \\ 0& 0}\ ,\\
Q_2&=\MAT{ 1 & 0\\
0& 0}\ ,\quad \tilde Q_2=\MAT{m_1(\mu_1-\mu_2)+m_2(\mu_1+\mu_2) & 0\\ 0& 0}\ . }
The other Higgs
vacuum is obtained by swapping $\Phi_1\leftrightarrow\Phi_2$ along with
\SP{ Q_1&=\MAT{0 & 0\\ 1 & 0}\ ,\quad \tilde Q_1=\MAT{0
& m_1(\mu_1-\mu_2)+m_2(\mu_1+\mu_2) \\ 0& 0}\ ,\\
Q_2&=\MAT{ 0 & 0\\
0& 1}\ ,\quad \tilde Q_2=\MAT{0 & 0\\ 0& m_1(-\mu_1+\mu_2)+m_2(\mu_1+\mu_2)}\ .
} 
In total, therefore, 
there are 6 vacua with a mass gap: 4 confining and 2 Higgs.

There are two massless, or Coulomb, 
vacua each with an unbroken $\U(1)$ factor. For the first
\SP{\Phi_1&
=\tfrac{\mu_2m_2}{\mu_1+\mu_2}{\rm diag}\big(1,-1\big)\ ,\qquad\Phi_2=
\tfrac{\mu_1m_2}{\mu_1+\mu_2}{\rm diag}\big(-1,1\big)\ ,\\
Q_1&=\tilde Q_1=0 ,\quad Q_2=\MAT{1 &0 \\0 & 0}\ ,\quad \tilde
Q_2=\tfrac{4m_2\mu_1\mu_2}{\mu_1+\mu_2}\MAT{1 &0 \\ 0& 0}\ , } 
whilst for the second
\SP{\Phi_1&=\tfrac{\mu_2m_1}{\mu_1+\mu_2}{\rm diag}\big(1,-1\big)\ ,\qquad\Phi_2=
\tfrac{\mu_1m_1}{\mu_1+\mu_2}{\rm diag}\big(1,-1\big)\ ,\\
Q_1&=\MAT{0 &0 \\1 &0 }\ ,\quad \tilde
Q_1=\tfrac{4m_1\mu_1\mu_2}{\mu_1+\mu_2}\MAT{0 & 1\\ 0& 0}\ ,\quad Q_2=\tilde
Q_2=0\ .}

The analysis above holds for generic values of the
masses. However, for particular values of the masses
flat directions emerge and different vacua can be related. Of course
at this stage we emphasize that we are not taking account any of the
quantum effects. To start with, if $m_1$ or $m_2$ vanish then
the two Higgs vacua are related by a flat direction. For instance with
$m_2=0$ we have
\SP{
\Phi_1&=\Phi_2=\tfrac12m_1{\rm diag}\big(1,-1\big)\ ,\qquad
Q_1=\MAT{0 &0 \\ 1& 0}\ ,\\ \tilde Q_1&=\MAT{0 &m_1(\mu_1+\mu_2) \\ 0&0 }\
,\quad Q_2=\MAT{ 1&0 \\
0& 1}\ ,\quad \tilde Q_2=\MAT{x &0 \\ 0&x-m_1(\mu_1-\mu_2) }\ . } 
Here, $x$ parameterizes the flat
direction. In an analogous way, one of the massless vacua is 
related to the confining vacuum by a flat direction.

\section{The Exact Superpotential}

Having established in the last section a qualitative picture based on
the tree-level superpotential of the theory in four dimensions, we can now
investigate the phase structure exactly following 
Ref.~\cite{DHK1}. As we alluded to in
the introduction, the superpotential is precisely one of the
Hamiltonians in the space of action
variables of a complexified integrable system. For our theory the
latter is a
certain symplectic reduction of the 
spin generalization of the elliptic Calogero-Moser
system. We now construct it for the general $\SU(N)^k$ theory. In this
case it describes the motion of $N$ particles in one dimension
with positions $X_a$ and momenta $p_a$.
Each particle carries a ``spin'' in the form of 
a $k\times k$ matrix with elements $\J^a_{ij}$. The basic
Hamiltonian of the system is\footnote{We have written the following in
terms of spins $\J^a_{ij}$ which is slightly different
but equivalent to the way the system was written in \cite{DHK1} in
terms of the spins $\S^i_{ab}$. The relation
between the two representations can be determined from Eq.~\eqref{jtos}.}
\SP{
H_0&=\sum_ap_a^2+\sum_{a\neq
b}\sum_{ij}\J^a_{ji}\J^b_{ij}
\frac{\sigma(X_{ab}+z_{ji})}
{\sigma(X_{ab})\sigma(z_{ji})}\big(\zeta(X_{ab}+z_{ji})-\zeta(X_{ab})\big)\\
&-\tfrac12\sum_{i\neq
j}\Big[\sum_a\J^a_{ij}\J^a_{ji}-Nm_im_j\Big]
\Big(\wp(z_{ij})-\zeta(z_{ij})^2\Big)\ .
\label{ham}
}
Here, $\wp(z)$ is the Weierstrass function and 
$\sigma(z)$ and $\zeta(z)$ are its cousins
defined on the torus with half-periods $\omega_1=i\pi$ and
$\omega_2=i\pi\tau$ (so of complex structure $\tau$) \cite{WW}. In the above,
the separation between the particles is given by
$X_{ab}\equiv X_a-X_b$ while $z_{ij}\equiv z_i-z_j$ are
``inhomogeneities'', $k-1$ external parameters (since only the
differences matter). In our application, the $k$ independent complex coupling
constants $\tau_i$ of each of the $\SU(N)$ factors of the gauge group are
associated to the $k$ independent parameters $\{\tau,z_i\}$ in the
following way. Firstly we order the $z_i$ so that $0\leq{\rm Re}\,z_i\leq{\rm
Re}\,z_{i+1}\leq2\pi{\rm Im}\,\tau$. Then
\EQ{
\tau_i=i\frac{z_{i+1}-z_i}{2\pi}\qquad i=1,\ldots,k-1\ ,\qquad
\tau_k=i\frac{z_1-z_k}{2\pi}+\tau\ .
}

To define the dynamical system the dynamical variables have the
non-vanishing Poisson brackets \cite{Krichever:1994vg}
\EQ{
\{X_a,p_b\}=\delta_{ab}\ ,\qquad\{\J^a_{ij},\J^b_{kl}\}=\delta_{ab}\big(
\delta_{jk}\J^a_{il}-\delta_{il}\J^a_{kj}\big)\ .
\label{pbs}
}
In fact, in the application to gauge theory, 
the spins are not arbitrary $k\times k$ matrices, rather they
have rank one and so we can define them in terms of new variables
$Q_{ai}$ and $\tilde Q_{ia}$:
\EQ{
\J^a_{ij}=\tilde Q_{ia}Q_{aj}\ . 
\label{jtos}
} 
If we take all the inhomogeneities $z_i$
equal, then \eqref{ham} simplifies to
\EQ{
H_0=\sum_ap_a^2-\sum_{a\neq
b}{\rm Tr}\big(\J^a\J^b\big)\wp(X_{ab})\ ,
\label{hams}
}
the dynamical system analysed in
Ref.~\cite{Krichever:1994vg}. The system is completely integrable,
even when the $z_i$ are arbitrary,
so there exists a basis of action-angle variables for which the
Hamiltonian \eqref{ham} is but one of a set of action variables.

For the application to gauge theory, 
we have to impose additional conditions on the
spins. The reduction can be defined as a symplectic quotient by the
abelian symmetries
\EQ{
Q_{ai}\to e^{\phi_a}Q_{ai}e^{\psi_i}\ ,\quad \tilde Q_{ia}\to e^{-\psi_i}\tilde
Q_{ia}e^{-\phi_a}\ .
\label{wee}
}
In all there are $N+k-1$ independent symmetries. Taking the symplectic
quotient involves imposing the momentum map constraints:
\EQ{
\sum_aQ_{ai}\tilde Q_{ia}=Nm_i\ ,\qquad\sum_iQ_{ai}\tilde Q_{ia}=0\ ,
\label{mommap}
}
along with an ordinary quotient by the symmetries \eqref{wee}.
Notice that hypermultiplet masses $m_i$ enter via \eqref{mommap}.
In \eqref{ham} we note that the centre-of-mass motion is completely
trivial and so we set $\sum_ap_a=\sum_a X_a=0$. Once this has been done
the phase space (after the symplectic quotient of the spins) has the
dimension $2k(N-1)$: precisely the complex dimension of the Coulomb
branch of the compactified $\SU(N)^k$ theory.
So the complexified phase is identified with the Coulomb branch of the
compactified theory. This space is actually a hyper-K\"ahler manifold
with a chosen complex structure. This is clear in the
formulation of the integrable system
as a Hitchin system \cite{Kapustin:1998xn,DHK1,hitchin} which has the
form of an infinite-dimensional hyper-K\"ahler quotient \cite{Hitchin:1986ea}.
In this context, the symplectic form of the dynamical system is identified
with the closed $(2,0)$ form with respect to the chosen
complex structure. 

The remaining action variables can be extracted from the Lax operator
described in \cite{DHK1}. Of particular importance for us is the basic
Hamiltonian \eqref{ham} along with the following others
\SP{
H_i&=2\sum_ap_a\J^a_{ii}-2\sum_{a\neq
b}\sum_{j(\neq i)}\J^a_{ji}\J^b_{ij}
\frac{\sigma(X_{ab}+z_{ji})}
{\sigma(X_{ab})\sigma(z_{ji})}
+2\sum_{j(\neq
i)}\Big[\sum_a\J^a_{ij}\J^a_{ji}-Nm_im_j\Big]
\zeta(z_{ij})\ .
\label{ham2}
}
of which only $k-1$ are independent since $\sum_{i=1}^kH_i=0$.

The action variables, or Hamiltonians, parameterize the Coulomb branch
of the four-dimensional theory prior to compactification. In
particular, the $k$ independent Hamiltonians $H_0$ along with 
$H_i$ are identified with the subspace of quadratic condensates ${\rm
Tr}\,\Phi_i^2$. In \cite{DHK1}, we identified the unique combination of
Hamiltonians corresponding to the diagonal combination:
\EQ{
\sum_{i=1}^k{\rm
Tr}\,\Phi_i^2=k\,H^*\ ,
}
where 
\EQ{
H^*=H_0-{1\over k}\sum_{i\neq l}\zeta(z_{il})H_i\ .
\label{defhstar}
}
The fact that there is a non-trivial function multiplying the $H_i$ is
required in order that $H^*$ has the appropriate modular properties.
The superpotential in the three-dimensional compactification 
corresponding to an arbitrary $\N=1$ mass deformation of
the theory is then identified with a particular linear combination 
of the Hamiltonians:
\EQ{
W=\frac1{g^2}\sum_{i=1}^k\mu_i{\rm Tr}\,\Phi^2_i=
\frac 1{g^2}\Big(\lambda_0(\mu)H^*+\sum_{i=1}^k\lambda_i(\mu)H_i\Big)\ .
}
for quantities $\{\lambda_0,\lambda_i\}$ depending linearly on the $\mu_i$.

In the case of an $\SU(2)\times\SU(2)$ quiver, we can be 
more explicit. Firstly regarding the symplectic
reduction on the spins. Solving the moment map conditions
\eqref{mommap} and fixing the symmetries \eqref{wee} can be achieved,
for instance, by
parameterizing them with two variables $\{x,y\}$ such that
\EQ{
Q_{ai}=\MAT{1&m_2-y\\ e^{-x}\frac{m_1-y}{y-m_2}&m_2+y}\ ,\qquad
\tilde Q_{ia}=\MAT{y+m_1&e^x(y-m_2)\\ 1 & 1}\ .
}
The Poisson bracket that one derives from \eqref{pbs} is then simply
$\{x,y\}=1$.

Once this has been done, the dynamical system 
has a four-dimensional phase space
parameterized by $X\equiv X_1-X_2$, $p=\tfrac12(p_1-p_2)$, $x$ and $y$
with non-trivial Poisson brackets
\EQ{
\{X,p\}=1\ ,\qquad \{x,y\}=1\ .
}
The two Hamiltonians \eqref{ham} and \eqref{ham2} are
\AL{
H_0&=2p^2+2(2y^2-m_1^2-m_2^2)\wp(X)-
        2e^x(y^2-m_2^2)\frac{\sigma(X-z)}{\sigma(X) \sigma(z)} \big(
            \zeta(X-z)-\zeta(X) \big)\notag \\
        & +2e^{-x}(y^2-m_1^2) \frac{\sigma(X+z)}{\sigma(X) \sigma(z)}
            \big(\zeta(X+z)-\zeta(X)\big) +2y^2
(\wp(z)-\zeta(z)^2)\ ,\\
H_1&=4py-4y^2 \zeta(z)+ 2e^x(y^2-m_2^2)\frac{\sigma(X-z)}{\sigma(X)
\sigma(z)} +2e^{-x}(y^2-m_1^2)
\frac{\sigma(X+z)}{\sigma(X) \sigma(z)}\ , 
}
where $z\equiv z_{12}$. It is straightforward to check that $H_0$ and
$H_1$ Poisson-commute.

In the case of $\SU(2)\times\SU(2)$ we can uniquely identify the
relation between the gauge invariants operators ${\rm Tr}
\,\Phi_i^2$, $i=1,2$ and the Hamiltonians. Firstly, as in \eqref{defhstar}
there is a unique combination which has the
required properties to be identified with the
average combination:
\EQ{
\tfrac12{\rm Tr}\,\big(\Phi_1^2+\Phi_2^2\big)\equiv
H^*=H_0-\zeta(z)H_1\ .
}
whilst the quantity $H_1$ is identified with the difference
\EQ{
{\rm Tr}\,\big(\Phi_2^2-\Phi_1^2\big)\equiv H_1\ .
} 
It follows that
\EQ{
{\rm Tr}\,\Phi_1^2=H^*-\tfrac12 H_1\ ,\qquad
{\rm Tr}\,\Phi_2^2=H^*+\tfrac12 H_1\ . 
} 
We can now identify the general $\N=1^*$
deformation of the superpotential with the following linear combination of the
action variables:
\EQ{ W=\frac1{g^2}\big(\mu_1{\rm Tr}\,\Phi_1^2+\mu_2{\rm
Tr}_2\,\Phi_2^2\big)
=\frac1{g^2}(\mu_1+\mu_2)\tilde H\ , 
\label{pots}
} 
where 
\SP{
\tilde H&=H^*+\tfrac12\beta H_1
=2p^2+4\alpha p y +2e^x
(y^2-m_2^2)\tilde\phi(X)
     +2e^{-x}(y^2-m_1^2)\phi(X)\\
&+2(2y^2-m_1^2-m_2^2) \wp(X)
+2y^2 (\wp(z)-\zeta(z)^2-2\alpha\zeta(z))\ ,
\label{defht}
}
where we have defined the constant
\EQ{
\alpha=-\zeta(z)+\tfrac12\beta\ ,
\qquad\beta=\frac{\mu_2-\mu_1}{\mu_1+\mu_2} 
}
along with the functions
\SP{
     \phi(X) & = \frac{\sigma(X+z)}{\sigma(X) \sigma(z)} \left(
            \zeta(X+z)-\zeta(X)+\alpha \right)\ , \\
    \tilde\phi(X) & =\frac{\sigma(X-z)}{\sigma(X) \sigma(-z)} \left(
            \zeta(X-z)-\zeta(X)-\alpha \right)\ .
}

\section{The Exact Phase Structure}

Supersymmetric vacua are obtained by extremizing the
superpotential \eqref{pots}. One obtains the
equations
\AL{
\frac{\partial \tilde H}{\partial p} &= 4p+4\alpha y=0\ ,\label{cpa}\\
  \frac{\partial\tilde H}{\partial y} &= 4y\big\{2\wp(X)+\wp(z)
            +e^x\tilde\phi(X)+e^{-x}\phi(X)\big\}=0\ ,\label{cpb}\\
 \frac{\partial\tilde H}{\partial x} &= 2e^x(y^2-m_2^2)\tilde\phi(X)
            -2e^{-x}(y^2-m_1^2)\phi(X)=0\ ,\label{cpc}\\
  \frac{\partial\tilde H}{\partial X} &= 2(2y^2-m_1^2-m_2^2) \wp'(X)
            +2e^x(y^2-m_2^2) \tilde\phi'(X)
            +2e^{-x}(y^2-m_1^2)\phi'(X)=0\ .\label{cpd}
}

We now begin by solving \eqref{cpa} for $p$:
\EQ{
p=-\alpha y\ .
\label{sola}
}
One branch of solutions is then obtained by solving \eqref{cpb} with
$y=0$. It then follows that there are two solutions of \eqref{cpc},
which we label by $n_1=1,2$, for which
\EQ{
e^x=(-1)^{n_1} \frac{m_1}{m_2} \sqrt{\frac{\phi(X)}{\tilde\phi(X)}}\ .
\label{solb} 
} 
Using standard elliptic function identities, along with
\eqref{sola} and \eqref{solb}, the final equation \eqref{cpd} can be
recast in the form
\EQ{
\wp'(X) \Big(m_1^2 +(-1)^{n_1} m_1m_2\gamma\big(\phi\tilde\phi\big)^{-1/2}
+m_2^2\Big)=0\ , 
\label{fineq}
}
where we have defined the quantity
\EQ{
\gamma=2\wp(X)+\wp(z)-\frac{\beta^2}{4}\ . 
} 
For later use, one can show, again using standard elliptic function
identities, that
\EQ{
\phi(X)\tilde\phi(X)=\wp^2(X)+\tfrac14\beta^2\wp(z)+\wp^2(z)+\wp(X)\wp(z)-
\tfrac14\beta^2\wp(X)+\tfrac12\beta\wp'(z)-\tfrac14g_2\ ,
}
from which one deduces
\EQ{
\gamma^2-4\phi\tilde\phi=g_2-3\wp^2(z)-2\beta\wp'(z)-\tfrac32\beta^2\wp(z)
+\tfrac1{16}\beta^4\ .
\label{idgam}
}
As a consequence the left-hand side is independent of $X$. 

For generic
masses the solution to \eqref{fineq} is $\wp'(X)=0$, {\it
 i.e.\/}~$X$ is a half-period
\EQ{
X\in\{\omega_1,\ \omega_2,\ \omega_1+\omega_2\}\ , 
\label{valx}
} 
which we label 
$X_c=i\pi,i\pi\tau,i\pi(\tau+1)$, $c=1,2,3$.

In order to assess whether these six vacua are massive or massless, we
compute the Hessian:
\SP{ 
{\rm
Det}\left[\frac{\partial^2\tilde H}{\partial x_i
\partial x_j}\right]&= \frac1{m_2^2\phi\tilde\phi}\Big((-1)^{n_1}
m_1m_2\gamma+(m_1^2+m_2^2) (\phi
\tilde\phi)^{1/2}\Big)\\
&\times \Big(2\phi\tilde\phi\wp''(X)\Big[(-1)^{n_1} m_1m_2\gamma+(m_1^2+m_2^2)
(\phi
\tilde\phi)^{1/2}
\Big]\\
&\qquad -(-1)^{n_1} m_1m_2(\gamma^2-4\phi\tilde\phi) \wp'(X) \Big)\Big|_{X=X_c}\ . 
}
It can be shown that the above is generically
non-zero so that all six vacua are massive.
The values of the condensates in these six massive vacua are
\SP{ 
&{\rm Tr}\,\,\Phi_i^2= -2(m_1^2+m_2^2)\wp(X)\\
&-4(-1)^{n_1}\frac{m_1
m_2}{(\phi\tilde\phi)^{1/2}} 
\Big(\phi
\tilde\phi+\tfrac12
(\beta-(-1)^i)\big(\beta(\wp(z)-\wp(X))+\wp'(z)\big)\Big)\Big|_{X=X_c}\ .
\label{massca}}

These six vacua are precisely the vacua found in \cite{DHK1} 
for general $k$ and $N$. It is tempting to
identify them with the six massive vacua, two Higgs and four
confining, that we found in Section 2 and this turns out to be
correct. In order to pin down the relation, consider the
semi-classical expansion of the condensates in each of the vacua as
described in \cite{Hollowood:2002ax}.
The expansions we need can be deduced from the following expansions of
the (quasi-)elliptic functions
\SP{
\wp(X)&=\frac1{12}+\frac{e^{-X}}{(1-e^{-X})^2}+\sum_{n=1}^\infty\Bigg\{
\frac{e^{-X}q^n}{(1-e^{-X}q^n)^2}+\frac{e^Xq^n}{(1-e^Xq^n)^2}-
\frac{2q^n}{(1-q^n)^2}\Bigg\}\ ,\\
\sigma(X)&=e^{\zeta(i\pi)X^2/(2i\pi)}(e^{X/2}-e^{-X/2})\prod_{n=1}^\infty
\frac{1-q^n(e^X-e^{-X})+q^{2n}}{(1-q^n)^2}\ ,\\
\zeta(X)&=X\frac{\zeta(i\pi)}{i\pi}+\frac12\coth(X/2)-\sum_{n=1}^\infty
\frac{q^n(e^X+e^{-X})}{1-q^n(e^X+e^{-X})+q^{2n}}
\ .
}
The condensates can be written in terms of the complex couplings of
each gauge group factor:
\EQ{
q_1=e^{2\pi i\tau_1}=e^{-z}\ ,\qquad q_2=e^{2\pi i\tau_2}=qe^{z}\ ,
}
where $q=e^{2\pi i\tau}$. It is easy to see that the condensates have
an expansion in terms of the quantities
\EQ{
e^{-X}q^n\ ,\quad e^Xq^{n+1}\ ,\quad q^n\ ,\quad e^{-z}q^n\ ,\quad
e^zq^{n+1}\ ,
}
with $n=0,1,2,\ldots$.
Given the values for $X$ in \eqref{valx}, it is clear that the vacua
with $X=i\pi$ have an expansion in integer powers of $q_1$ and
$q_2$. Hence, the two vacua with $X=i\pi$ are identified with the
Higgs vacua in which the condensates have a conventional
semi-classical instanton expansion
in integer powers of $q_1$ and $q_2$. The vacua with $X=i\pi\tau$ or
$i\pi(\tau+1)$ have an expansion which includes powers of the fractional
instanton factor $q^{1/2}$. This is characteristic of a confining
vacuum. Hence, we identify the four vacua with these values of $X$ and
$n_1=1,2$ with the four confining vacua identified in
Section 2. 

Now we return to the equations for the vacua \eqref{cpa}-\eqref{cpd}
and choose a different branch of solutions obtained by solving
\eqref{cpb} with
\EQ{
e^x=\frac{-\gamma +(-1)^{n_2}\sqrt{\gamma^2 -4 \phi\tilde\phi}}{2\tilde\phi}\ ,
\label{masslx} }
rather than $y=0$. There are two solutions of this type labelled by 
$n_2=1,2$. Then \eqref{cpc} is solved for $y$ giving
\EQ{
y=\sqrt{\frac{m_2^2e^x\tilde\phi-m_1^2e^{-x}\phi}
{e^x\tilde\phi-e^{-x}\phi}}\ . 
\label{massly} 
}
Choosing the opposite sign for $y$ can be shown to lead 
to an equivalent solution due to the presence of discrete symmetries
which we have hitherto ignored. In particular the values of the
condensates will not depend on it.

The final equation \eqref{cpd} becomes
\SP{ 
(m_1^2-m_2^2)\frac{\partial
\sqrt{\gamma^2-4\phi\tilde\phi}}{\partial X}=0 
} 
which is identically zero for
all values of $X$ since the combination \eqref{idgam} is independent
of $X$. 

The two solutions are obviously massless vacua since 
each corresponds to a line of critical
points parameterized by $X$.
The values of the condensates in these two massless vacua are 
\SP{
{\rm Tr}\,\,\Phi_i^2=
&(m_1^2+m_2^2)\big(\wp(z)-\tfrac12(-1)^i\beta+\tfrac14\beta^2\big)
\\
&+(-1)^{n_2}\frac{m_1^2-m_2^2}{\sqrt{\gamma^2-4\phi\tilde\phi}}
\Big(\gamma^2-4\phi\tilde\phi+\tfrac12
\beta(\beta-(-1)^i)\big(3\wp(z)+2\wp'(z)-\tfrac14\beta^2\big)\Big) 
\label{masslca}
}

The discussion of the vacuum structure above has been established in
the case where the masses $\{m_i\}$ and $\{\mu_i\}$ are generic. For
special values the vacua can merge. First of all, if $X$ equals a half
period and $y$ in \eqref{massly} equals 0, which requires
\EQ{
m_2^2 e^x \tilde\phi-m_1^2e^{-x}
\phi=0\ ,
}
where $x$ is given by \eqref{masslx}, then a massless vacuum meets
what was once one of the massive vacua. Solving these
equations leads to a condition on the ratio of the hypermultiplet
masses $m_1/m_2$. In this way
either of the massless vacua can meet any of the 6 massive vacua
at 12 special values for $m_1/m_2$:
\EQ{
\frac{m_1}{m_2}=(-1)^{n_1}\frac{-\gamma+(-1)^{n_2}\sqrt{\gamma^2-4\phi\tilde\phi}}
{2\sqrt{\phi\tilde\phi}} } with $X=X_c$, $c=1,2,3$. 
Finally the two massless vacua merge
together when
\EQ{
\gamma^2-4\phi\tilde\phi=g_2-3\wp^2(z)-2\beta\wp'(z)-3\beta^2\wp(z)/2+\beta^4/16=0\
. }

\section{Discussion}

We have calculated the exact phase structure and the condensates of
the two adjoint-valued scalar fields in the mass deformed
$\SU(2)\times\SU(2)$ finite quiver theory. The strategy involved
compactifying the theory on a circle of finite radius so that the
low-energy degrees-of-freedom are all scalar. However, the values
calculated remain valid in the decompactification limit. In this way,
we were able to show how the exact structure of vacua matches the one
deduced from an analysis of the tree-level superpotential in the
four-dimensional theory. It would be interesting to extend our
analysis to the general $\SU(N)^k$ quiver theories and also to
consider the solution of these mass-deformed theory using the matrix
model formalism developed by Dijkgraaf and Vafa \cite{Dijkgraaf:2002dh}
also applied to the ${\cal N}=1^*$ theory in
\cite{Dorey:2002tj,Dorey:2002jc}. 

\vspace{2cm}

We would like to thank Nick Dorey and Prem Kumar for many useful
conversations on the mass deformed quiver theories.

\end{document}